\documentclass[lettersize,journal]{IEEEtran}
\pdfoutput=1
\usepackage{amsmath,amsfonts}
\usepackage{amssymb}
\usepackage[ruled,linesnumbered]{algorithm2e}
\usepackage{array}
\usepackage{algpseudocode}
\usepackage[caption=false,font=normalsize,labelfont=sf,textfont=sf]{subfig}
\usepackage{textcomp}
\usepackage{stfloats}
\usepackage{url}
\usepackage{verbatim}
\usepackage{graphicx}
\usepackage{cite}
\usepackage{hyperref}
\hypersetup{hidelinks,
	colorlinks=true,
	allcolors=black,
	pdfstartview=Fit,
	breaklinks=true}
\usepackage{stfloats}
\usepackage{amsmath}
\usepackage{booktabs}
\usepackage{comment}

\makeatother
\usepackage{ragged2e}
\usepackage[explicit]{titlesec}
\usepackage[caption=false, font=normalsize, labelfont=sf, textfont=sf]{subfig}
\UseRawInputEncoding
\begin{document}
\title{Towards Decentralized Task Offloading and  Resource Allocation in User-Centric MEC}

\author{Langtian Qin, Hancheng Lu,~\IEEEmembership{Senior Member,~IEEE}, Yuang Chen, Baolin Chong and Feng Wu,~\IEEEmembership{Fellow,~IEEE}
\IEEEcompsocitemizethanks{\IEEEcompsocthanksitem L. Qin, H. Lu, Y. Chen, B. Chong and F. Wu are with the Department of Electronic Engineering and Information Science, University of Science and Technology of China, Hefei 230027, China. (E-mail: qlt315@mail.ustc.edu.cn; hclu@ustc.edu.cn; yuangchen21@mail.ustc.edu.cn; chongbaolin@mail.ustc.edu.cn; fengwu@ustc.edu.cn)  \protect\\

 }
}

\maketitle
\begin{abstract}
In the traditional cellular-based mobile edge computing (MEC), users at the edge of the cell are prone to suffer severe inter-cell interference and signal attenuation, leading to low throughput even transmission interruptions. Such edge effect severely obstructs offloading of tasks to MEC servers. To address this issue, we propose user-centric mobile edge computing (UCMEC), a novel MEC architecture integrating user-centric transmission, which can ensure high throughput and reliable communication for task offloading. Then, we formulate an optimization problem with joint consideration of task offloading, power control, and computing resource allocation in UCMEC, aiming at obtaining the optimal performance in terms of long-term average total delay. To solve the intractable problem, we propose two decentralized joint optimization schemes based on multi-agent deep reinforcement learning (MADRL) and convex optimization, which consider both cooperation and non-cooperation among network nodes. Simulation results demonstrate that the proposed schemes in UCMEC can significantly improve the uplink transmission rate by at most 343.56\% and reduce the long-term average total delay by at most 45.57\% compared to traditional cellular-based MEC.
\end{abstract}


\section{Introduction}
By deploying servers near the user's side, such as the base station (BS) in the wireless access network, mobile edge computing (MEC) enables users with limited computing capacity to offload tasks for faster processing\cite{Shi2016}. Since users need to offload tasks to BS through wireless transmission, MEC services are closely related to and coupled with the transmission performance of wireless networks. With the rapid development of mobile internet, the rise of new applications such as virtual reality (VR) and augmented reality (AR) places higher demands on transmission throughput and communication reliability of the wireless networks with MEC. However, in the traditional MEC architecture based on cellular networks, users at the edge of the cell or with obvious shadow fading will suffer severe inter-cell interference or signal attenuation, and the wireless transmission rates will be considerably reduced. When such users request MEC services (such as VR video rendering and streaming, etc.), large signal interference and poor wireless environment will significantly increase the probability of transmission interruptions or offloading failure\cite{Mukherjee2020}. Consequently, those users have to process tasks locally, leading to a substantial reduction in quality of service (QoS) and overhead increase. Thus, the network-centric MEC based on cellular networks fails to meet the transmission and computing requirements of some users when running delay-sensitive or computing-intensive applications\cite{Qin2023}.

Through extensively deploying access points (APs) and organizing networks around users' specific conditions and requirements, user-centric network (UCN) breaks the concept of "cell" in cellular networks, which can effectively reduce signal interference and improve transmission rate\cite{Ammar2022}. Inspired by the advantages of UCN, we propose user-centric mobile edge computing (UCMEC), a novel MEC architecture integrating user-centric transmission, which can ensure high throughput and reliable communication and computing services for all users. In UCMEC, each user will be assigned a set of APs, i.e., an AP cluster dynamically based on their geographic location and network environment. When a user need to offload tasks, all APs within its AP cluster will provide collaborative uplink transmission service. After receiving the user data, the AP cluster forward the signal to the central processing unit (CPU) integrated with MEC servers through wireless fronthaul. The CPU then perform signal decoding based on multiple copies of data sent by AP cluster and allocate computing resource to perform task processing.

However, the combination of UCN and MEC is non-trivial, since the AP collaborative uplink transmission and the wireless fronthaul transmission will make the task offloading and resource allocation in UCMEC more challenging then cellular-based MEC. Firstly, task offloading in uplink transmission via AP cluster leads to increased transmission interference among users in the system. To mitigate this, users must make carefully decisions regarding the uplink transmission power and the necessity of local processing to avoid excessive transmission interference and its associated overhead\cite{Hancheng2022}. Secondly, since AP clusters need to transmit offloaded task data to the CPU through wireless fronthaul, task offloading decisions will also impact the competition among APs for wireless fronthaul transmission resource. If too many APs utilize fronthaul for transmission, it results in heightened data volume and interference, ultimately leading to increased fronthaul transmission delay\cite{Ibrahim2022}. Thirdly, users must be prudent in selecting the appropriate CPU for task offloading, as this choice significantly affects both the uplink transmission and wireless fronthaul transmission process of the AP cluster\cite{UCNsurvey}. In addition, the task offloading decisions are coupled with the computing resource allocation of multiple CPUs with limited MEC capacities, as well as the power control of users. Hence, the joint optimization of task offloading and resource allocation becomes imperative for efficient system performance in UCMEC.

To address the aforementioned challenges, in this paper, we propose efficient task offloading and resource allocation schemes in UCMEC. Specifically, to minimize the long-term average total delay of the system, we jointly optimize task offloading, power control, and computing resource allocation.  Considering the dynamic of the user demand, wireless environment as well as the actual algorithm deployment, we design two decentralized optimization schemes based on convex optimization and multi-agent deep reinforcement learning (MADRL) for two cases: case with node collaboration and case without node collaboration respectively. The main contributions of this paper are summarized as follows:

\begin{itemize}
    \item {We proposed UCMEC, a novel MEC architecture that incorporating user-centric transmission and MEC task processing. By analyzing the delay of uplink transmission, fronthaul transmission, and task processing, we formulate a long-term average delay minimization problem by jointly optimizing task offloading, power control, and computing resource allocation.}

    \item {To address the intractable dynamic problem, we propose the decentralized joint optimization scheme based on MADRL and convex optimization. Each user will act as an agent to obtain task offloading and power control strategy through MADRL, and the computing resource allocation problem will be solved by CPUs via convex optimization.}

    \item {Multi-agent proximal policy optimization (MAPPO)-based MADRL is adopted for the case that nodes are willing to collaboration, where users make decisions with information sharing and CPUs are responsible for training. For the case without node collaboration, we adopt independent proximal policy optimization (IPPO)-based MADRL where each selfish user makes independent decisions and training without information sharing.}
\end{itemize}
Extensive simulations are conducted to demonstrate the effectiveness of the proposed schemes. The results show that our decentralized optimization schemes in UCMEC can significantly improve the transmission rate and reduce the long-term average total delay compared to traditional cellular-based MEC. 

The rest of the paper is organized as follows: Section II presents the system model and problem formulation. Section III outlines the non-cooperative decentralized optimization scheme. Section IV describes the decentralized optimization scheme based on node collaboration. The evaluation of the proposed optimization schemes is presented in Section V. Section VI discusses related work, followed by the conclusion in Section VII. The list of key notations is shown in Table I.

\begin{table*}[tbp]
\centering
\caption{LIST OF KEY NOTATIONS}
	\renewcommand\arraystretch{1.2}
	\resizebox{\textwidth}{!}{
\begin{tabular}{|l|l|}
\hline
\textbf{Notation} & \textbf{Definition}  \\ \hline
 $M$ / $N$ / $K$ & Number of users / APs / CPUs                  \\ \hline
 $\rho$  & Number of antennas of APs \\ \hline
$D_m$ / $\rho_m$  & Data size / computing density of task of user $m$ \\ \hline
  $C^{l}_m$/ $C^{e}_k$    &  Computing capability of user $m$ / CPU $k$  \\ \hline
 $g^{a}_{mn}$  & Access channel between user $m$ and AP $n$  \\ \hline
 $\beta_{mn}$ / $h_{mn}$ / $PL_{mn}$ / $\mu_{mn}$  & Large-scale fading parameter / small-scale fading parameter / path loss /  shadow fading correction factor between user $m$ and AP $n$ \\ \hline

 $d^{a}_{mn}$ / $d^{f}_{mn}$   &  Distance between user $m$ and AP $n$  / distance between AP $n$ and CPU $k$ /  \\ \hline
 $\xi$ / $G_t$ / & Blockage density parameter / transmit gain of mmWave-fronthaul channel   \\ \hline

 $p_{nk}^{l}$ / $\alpha^{l}$  & Probability / Path-loss exponent  \\ \hline
 $p_{nk}^{n}$ / $\alpha^{n}$ & Probability / Path-loss exponent \\ \hline
 $B^f$ / $B^{a}$  & Bandwidth of access links / fronthaul links  \\ \hline
 $p_m^{p}$ / $p_m^{d}$ / $p_n^{f}$ & Transmit power for pilot transmission / offloading data of user $m$ / transmit power of AP $n$ \\ \hline
 $\omega_{mk}$ /  $c_{mk}$  & Offloading / computing resource allocation variable between user $n$ and CPU $k$              \\ \hline
  $r^{a}_{m}$ /  $r^{f}_{mk}$   & Uplink transmission rate of user $m$ / AP $n$    \\ \hline
 $\zeta^{a}_{m}$ /  $\zeta^{f}_{mk}$   & SINR of user $m$ / AP $n$  \\ \hline
 $T^l_{m}$  /$T^a_{m}$ / $T^f_{m}$ / $T^e_{m}$ / $T_{m}$ & Local processing / access transmission / fronthaul transmission / edge processing / total  delay of user $m$\\ \hline
\end{tabular}}
\end{table*}

\section{Related Work}
\textit{\textbf{Task Offloading and Resource Allocation in MEC:}} task offloading and resource allocation in MEC has been extensively studied
by previous work. To minimize the weighted age of information (AoI) of all the terrestrial user equipments, the authors in \cite{Zhen2023} jointly optimize the task scheduling, computing resource allocation, and unmanned aerial vehicle (UAV) trajectory based on the alternating optimization approach. In \cite{Apostolopoulos2023}, the authors formulate the optimal user task offloading to the available computing choices as a non-cooperative game of each user's satisfaction. By characterizing the secrecy-based throughput of wireless devices in its offloading transmission, the authors in \cite{Liping2023} jointly optimize the task offloading, secrecy provisioning, and offloading-transmission duration strategies to minimize the total energy consumption. The authors in \cite{Tong2022} propose an online task offloading and resource allocation approach for edge-cloud orchestrated computing, with the aim to minimize the average latency of tasks over time. An online learning framework LFSC is proposed in \cite{Ruiting2022}, which has the performance guarantee to guide task offloading in a small cell network. In \cite{Xiaoqian2023}, the authors propose a joint optimization framework for multi-dimensional resource constrained mobile crowdsensing systems by jointly controlling the data sensing, transmission and computation offloading schemes in the system. However, the above schemes for cellular-based edge computing are not be applicable to UCMEC directly, because users in UCMEC need to transmit task data to all APs in the AP cluster, and the AP cluster will transmit the data to the CPU for cooperative decoding and processing, instead of directly transmitting it to a specific base station for decoding and processing.

\textit{\textbf{Transmission Optimization in Cell-free Networks or UCN:}} With the rise of cell-free MIMO and UCN, many work analyze or optimize the transmission performance of users in cell-free network or UCN. In \cite{Haijun2022}, the authors jointly optimize the power control and fronthaul combining strategies to maximize the minimum signal to interference plus noise ratio (SINR). Also, a pilot assignment scheme based on inspection robot location is proposed to reduce pilot contamination. The authors in \cite{Demirhan2022} formulate the end-to-end data rate optimization problem accounting for the various practical aspects of the fronthaul and access links. a efficient joint beamforming and resource allocation solution is proposed. In \cite{Hancheng2022}, passive beamforming at reconfigurable intelligent surfaces (RIS) and power control at AP are jointly optimized  to maximize the energy efficiency of the RIS assisted user-centric network. The authors in \cite{Pinto2022}  study and compare the performance of coherent transmission and non-coherent transmission by formulating novel energy efficiency (EE) maximization problems for both strategies. the authors derive closed-form equations to find suboptimal solutions of both problems using a unified framework that combines successive convex approximation and the Dinkelbach algorithm. Two distributed downlink user scheduling, beamforming, and power control algorithms for user-centric network are proposed in \cite{Ammar2022}. The first algorithm is is implemented at the distributed units (DUs), while the second is implemented at the central units (CUs) controlling these DUs. Different from those work that focus on transmission optimization, we integrate MEC service into UCN, which will make problem analysis and solving more complex, as the transmission, task offloading and processing should be jointly considered.

\textit{\textbf{MEC with Cell-free Networks or UCN:}} Some work attempts to integrate MEC with the user-centric ideology. The authors in \cite{Sige2023} minimize the long-term delay for a given task duration under a price budget constraint. To address this problem, the authors develop a novel contextual sleeping bandit learning (CSBL) algorithm, which integrates contextual information and sleeping characteristics to accelerate learning convergence. Also, leverage Lyapunov optimization is adopted to deal with the price budget constraint. In \cite{Xiaoheng2021}, the authors consider energy consumption, delay, and price, and propose a joint optimization loading scheme to minimize the weighted cost of time delay, energy consumption, and price under the constraint of satisfying the advanced personalized needs of users. The authors in  \cite{Malong2021} study massive access in cell-free massive multi-input multi-output (MIMO)-based Internet of Things and solve the challenging active user detection (AUD) and channel estimation (CE) problems. In \cite{Malong2021}, cloud computing represents that the signals received by the AP are centrally processed by the CPU. For edge computing, the central processing is offloaded to some APs equipped with distributed processing units, so that AUD and CE can be executed in a distributed processing strategy. It should be noted that the ``user-centric'' in  \cite{Sige2023} and \cite{Xiaoheng2021} refers to the focus on the user's requirements, rather than the AP cluster-based wireless transmission method. Edge computing and cloud computing in \cite{Xiaoheng2021} refer to the signal processing methods of APs in UCN, instead of the offloaded tasks of users. Therefore, they are quite different from our paper. On the other hand, similar to this paper, the authors in \cite{Tilahun2022} propose a distributed solution approach based on a cooperative multi-agent reinforcement learning framework to solve resource allocation problems in MEC with cell-free MIMO. In \cite{Mukherjee2020}, the authors consider the cell-free (CF) massive MIMO framework with implementing MEC functionalities and use stochastic geometry and queueing theory to optimize the successful edge computing probability (SECP). However, \cite{Tilahun2022} considers only one CPU and assumes the ideal fronthaul. while in \cite{Mukherjee2020}, the author assumes that the MEC server is integrated on APs instead of the CPU, which is not practical because the signal is not decoded on the AP, therefore the tasks are hard to be processed directly in AP. At the same time, \cite{Mukherjee2020} only analyze the offloading performance and did not provide the optimal task offloading and resource allocation decisions.

\section{System Model}

\subsection{Network Model}
As shown in Fig. 1, we consider an uplink UCMEC system. There are $M$ single antenna users and $N$ multi-antenna APs, as well as $K$ CPUs with integrated MEC servers that are arbitrarily distributed over the coverage area. The user set, AP set and the CPU set are represented by $\mathcal{M}={1,2,..., m,..., M }$, $\mathcal{N}={1,2,..., n,..., N}$, and $ \mathcal{K}={1,2,..., k,..., K}$, respectively. The number of antennas for each AP is $\rho$, and each AP will communicate with the CPU through wireless fronthaul. APs are responsible for transmitting radio frequency (RF) signals, and channel estimation, while the CPUs are responsible for signal decoding and task processing\cite{Ammar2022}. Assuming the data transmission on the fronthaul and access channels are separated through frequency division \cite{Ammar2022TCOM}. The network operations in a time slotted fashion, where the duration of each time interval is equal to the coherent time $\tau_c$, and the operating period can be indexed by $ \mathcal {T}= {0, 1,.... T }$. Considering the channel block fading model, the channel remains unchanged in each time slot $t \in \mathcal {T}$, but varies in different time slots. In each time slot, user $m$ has a computing-intensive task, which can be represented by a tuple, i.e., $Task_m(t)= \{D_m(t), \rho_m (t)\}$, where $D_m(t)$ represents the data size (in bit) and $\rho_m (t)$ represents the computing density (in CPU cycles per bit), respectively. Assume that each task has a maximum tolerable delay $\tau_c$, users can choose to process the tasks locally or offload tasks to the MEC servers on the CPUs through APs. We assume that the local computing capacity of user $m$ in each time slot is $C_m^{\rm l}$ (in CPU-cycle frequency), while the computing capacity of CPU $k$ is $C_k^{{\rm e}}$

\subsection{Channel Model}

\subsubsection{Sub-6GHz-based Access Channel Model}
Since the environment around users includes static and dynamic objects that may completely block the LoS links between users and APs\cite{Ibrahim2022}. To ensure the reliable access communications between users and APs, we consider that access links operate on the sub-6GHz band. Assume $\boldsymbol g^{a}_{mn}(t) \in \mathbb{C}^{\rho \times 1}$ represents the channel between user $m$ and AP $n$ in time slot $t$, which remains unchanged during coherent time and satisfies the following equation:
\begin{equation}
    \boldsymbol g^{a}_{mn}(t) = \sqrt{\beta_{mn}(t)} \boldsymbol h_{mn}(t),
\end{equation}
where $\beta_{mn}(t)$ is the scalar coefficient of large-scale fading between user $m$ and AP $n$, $\boldsymbol  h_{mn}(t)$ is the small-scale fading coefficient between user $m $and AP $n$, and we have $\boldsymbol h_{mn}(t)\sim \mathcal{CN}(0,\boldsymbol I_{\rho})$. The small-scale fading is assumed to be static during each coherence interval, and change independently from one coherence interval to the next. The large-scale fading changes much more slowly, and stays constant for several coherence intervals depending on the user mobility. Similar to \cite{Ngo2017}, $\beta_{mn}(t)$ is given by:
\begin{equation}
    \beta_{mn}(t) = 10^{\frac{PL_{mn}(t)}{10}}10^{\frac{\sigma^s \mu_{mn}(t)}{10}},
\end{equation}
where $PL_{mn}(t)$ is the path-loss (in dB) between user $m$ and AP $n$ in time slot $t$, $10^{\frac{\sigma^s \mu_{mn}(t)}{10}}$ represents the shadow fading with a standard deviation of $ sigma^s$, $ mu_{mn}(t)$ is the shadow fading coefficient. For $PL_ {mn} (t) $, we use the following three slope path-loss model\cite{Buzzi2018}:

\begin{equation}
\begin{cases}
-L-35 {\rm log}_{10}(d_{mn}(t)),  & d_{mn}(t) > d_1\\
-L-10 {\rm log}_{10}(d_1^{1.5}d_{mn}^2(t)),  & d_0 < d_{mn}(t) \leq d_1 \\
-L-10 {\rm log}_{10}(d_1^{1.5}d_0^2), & d_{mn}(t) < d_0, 
\end{cases}
\end{equation}
where $d_{mn}(t)$ represents the distance (in km) between user $m$ and AP $n$ in time slot $t$, $d_0$and $d_1$ are the breakpoints of the three slope path-loss model, where $L$ satisfies the following equation:
\begin{equation}
\begin{aligned}
        L = &46.3 + 33.9{\rm log_{10}}(f) - 13.82{\rm log_{10}}(h_{a}) - [1.11{\rm log_{10}}(f) \\
        &-0.7]h_{u} + 1.56{\rm log_{10}}(f)-0.8,
\end{aligned}
\end{equation}
where $f$ represents the carrier frequency (in MHz), and $h^{ \rm a} $and $h^{ \rm u} $represent the antenna heights of the AP and the user, respectively. Due to the presence of obstacles between AP and user, by correcting the shadow fading, $ mu_{mn}(t)$ consists of two parts:
\begin{equation}
    \mu_{mn}(t) = \sqrt{\delta}\ddot{\mu}_n(t) + \sqrt{1-\delta}\ddot{\mu}_m(t),
\end{equation}
where $\dot{\mu}_n(t)$ and $\ddot{\mu}_m(t)$ follow the Gaussian distribution $ \mathcal {N} (0,1) $, and $0  \leq  \delta  \leq 1$ is a parameter.

\begin{figure}[tbp]
	\centering
	\includegraphics[angle=0,scale=0.3]{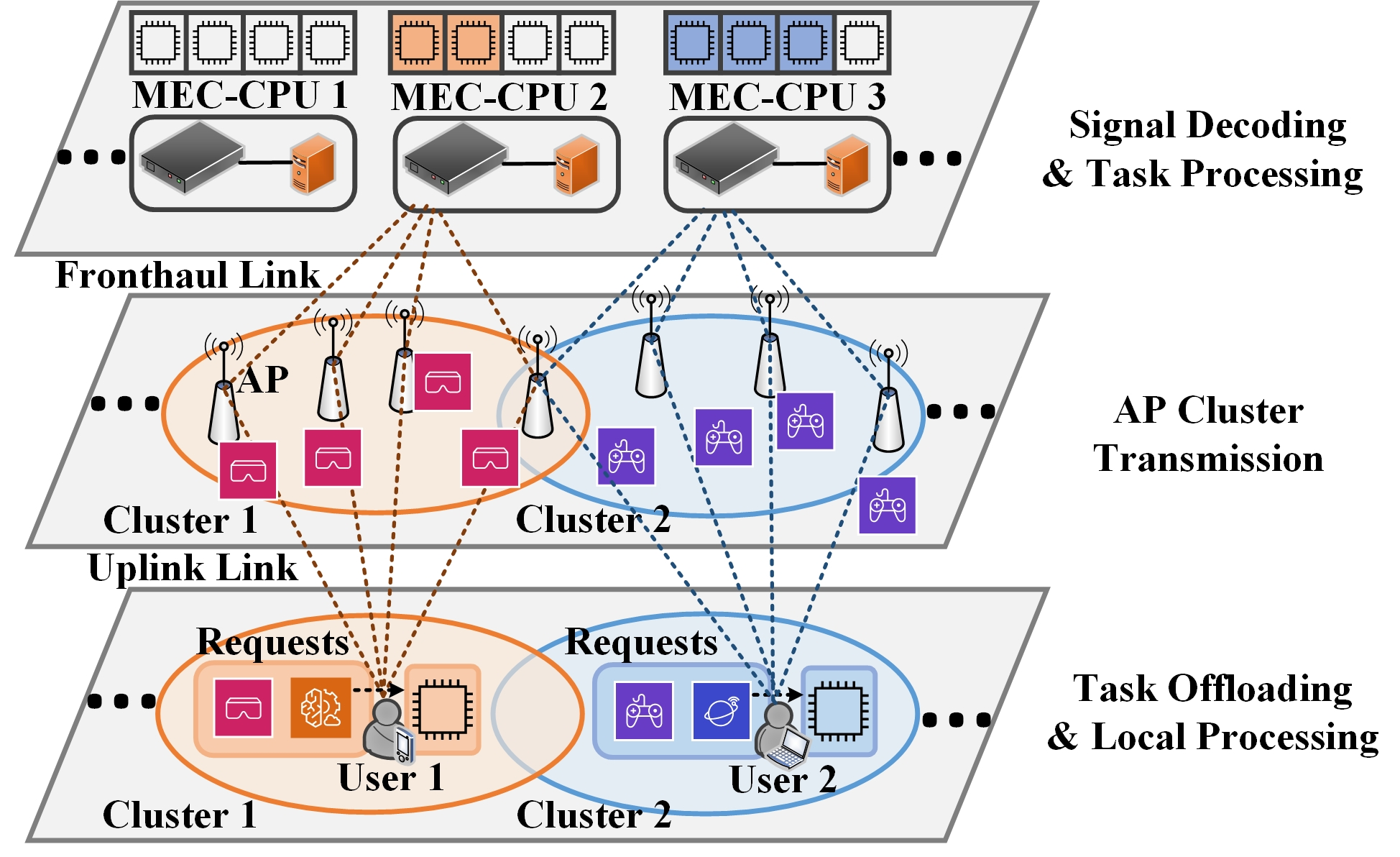}
	\caption{System architecture of Mmwave-Fronthaul UCMEC .}\label{fig:application}
\end{figure}

\subsubsection{MmWave-based Fronthaul Channel Model}
 MmWave-based transmission is adopted for the fronthaul due to the large available bandwidth and the high beamforming gains. Such high bandwidth can handle the immense amount of data transmission in the fronthaul of UCMEC. Nevertheless, a prerequisite for a reliable operation of the mmWave fronthaul network is to have LoS links between the CPUs and their associated APs. This is due to the fact that signal transmission in such band suffers from severe path-loss, especially under non line-of-sight (NLoS) transmissions, which in turn may lead to transmission breakdown\cite{Yazheng2021}. Similar to \cite{Tong2021,Kusaladharma2020}, we adopt the negative exponential function as the blockage probability model for mmWave-based fronthaul links, such that the probability of the link to be LOS or NLOS is a function of the distance between the AP and the corresponding serving CPU\cite{Andrews2017}. Assume that the distance between AP $n$ and CPU $k$ is $d^{f}_{nk}(t)$, then the probability that the mmWave-based fronthaul between AP $n$ and CPU $k$ at time slot $t$ is LOS or NLOS can be modeled as

\begin{equation}
    p^{l}_{nk} = e^{-\xi d_{nk}}, p^{n}_{nk} = 1 - e^{-\xi d_{nk}}
\end{equation}
respectively, where $\xi$ is the blockage density parameter\cite{Ibrahim2022}. In this paper, we consider the blockage effects by using the defined blockage probability function, and we assume the LOS probabilities between different mmWave links are independent\cite{Shi2019}. For the propagation characteristic in mm-wave bands, it is known that the small-scale fading in mm-wave channels is very weak, in contrast to that in microwave bands. Therefore, the small-scale fading can be neglected in mmWave band\cite{Guang2018,Delin2020}. Assume that the antenna arrays at APs and CPUs perform directional beamforming. The main lobe is directed towards the dominant propagation path and having less radiant energy in other directions. For tractability in the analysis, we adopt a sectorial antenna pattern\cite{Renzo2015}. Denote $\varphi$ as the main lobe beamwidth, $G^m$ and $G^s$ as the directivity gain of main and side lobes, respectively. Then the random antenna gain/interference $G_{nk}$ between the AP $n$ and the CPU $k$ has 3 patterns with different probabilities, which is given as

\begin{equation}
     G_{nk}=\left\{
\begin{array}{rcl}
\left(G^m\right)^2      &      & {\text{with prob.} (\frac{\varphi}{2\pi})^2},\\
G^m \times G^s     &      & {\text{with prob.} (\frac{2\varphi(2\pi-\varphi)}{(2\pi)^2})^2},\\
 \left(G^s\right)^2    &      & {\text{with prob.} (\frac{2\pi-\varphi}{2\pi})^2}.\\
\end{array} \right. 
\end{equation}

For the path-loss model, we assume the reference distance is 1 m, and the transmit power with propagation distance is attenuated by factor ${d^{f}_{nk}}^{-\alpha}$ where $\alpha$ is the path-loss exponent \cite{Ibrahim2022,Han2018}. When AP $n$ sends data to its associated CPU $k$, the SINR at CPU $k$ is given by

\begin{equation}
    \zeta^{f}_{nk}(t) = \frac{p^{f}_n G_{nk} {d^{f}_{nk}}^{-\alpha}} {I_{nk}(t) + {\sigma_{f}}^2 },
\end{equation}

where $P_n^{f}$ is the power of AP $n$ for fronthual transmission, and we have $\alpha = \alpha^{l}$ for LOS links while $\alpha = \alpha^{n}$ for NLOS links\cite{Gu2022}. ${\sigma_{f}}^2$ is the noise variance of the mmWave link. $I_{nk}(t) = \sum_{i \in \mathcal{N}} \sum_{j \in \mathcal{K}} g_{ij}^{f}(t)  p^{f}_i {d^{f}_{ij}}^{-\alpha} - g_{nk}^{f}(t) p_n^f {d^{f}_{nk}}^{-\alpha}$ represents the interference caused by other APs. Among them, $g_{ij}^{f}(t)$ is a binary variable, $g_{ij}^{f}(t)=1$ represents that at time slot $t$, AP $i$ will send data to CPU $j$, otherwise not. This variable is related to the task offloading decision and the AP clustering results. (specific relationships will be introduced in Sec. III. E).

\subsection{Channel Estimation}
To acquire channel state information (CSI),  each of the coherence intervals, i.e. coherence time and coherence bandwidth, is divided into two phases: uplink channel training to estimate the access channels between each of the User–AP pair, and uplink data transmission. Without loss of generality, we assume that the CPU has perfect CSI of fronthaul links at each time interval due to the fixed-distance fronthaul links\cite{Agheli2020}.

Similar to \cite{Chen2019}, we assume that at the beginning of each coherence time $\tau_{c}$, user $m$ needs to send a pilot sequence  $\boldsymbol {\lambda}_{m}(t) \in \mathbb{C}^{1 \times \tau_p}$, which has a length of $\tau_p$, where $||\boldsymbol \lambda_{m}(t)||^2=1$ and $\tau_p < \tau_c$. The underlying assumption is that $\tau_p \geq K$ to ensure the orthogonality of the pairwise pilot sequences\cite{Ngo2017}. Therefore, the received signal of AP $n$ at time slot $t$, i.e., $\boldsymbol Y_n(t) \in \mathbb{C}^{\rho \times \tau_p}(t)$ is \cite{Ammar2022}：
\begin{equation}
    \boldsymbol Y_n(t) = \sum_{m \in \mathcal{M}}{\sqrt{p_m^{p}}\boldsymbol g^{a}_{mn}(t)\boldsymbol \lambda_{m}(t)} + \boldsymbol Z_n(t),
\end{equation}
where $p_m^{p}$ is the transmit power of user $m$ when sending the pilot signal. Similar to \cite{Buzzi2017 ,Buzzi2018}, we assume that the user sends the pilot signal with the maximum transmit power $p^{max}$. $\boldsymbol Z_n(t)$ is the channel noise, where each element follows the i.i.d. complex Gaussian distribution $ \mathcal{CN} (0, {\sigma_a} ^ 2)$. After receiving the user's pilot signal, the AP will estimate the channel. In this paper, we adopt minimum mean-square error (MMSE) estimation\cite{Ammar2022}, the estimated channel at time slot $t$
$\boldsymbol{\hat{g}}^{a}_{mn}(t)$ can be given by
\begin{equation}
    \boldsymbol{\hat{g}}^{a}_{mn}(t) =\frac{\sqrt{\tau_p p^{max}} \beta_{mn}(t)\boldsymbol Y_n(t) \boldsymbol \lambda_m^{H}(t)}{\tau_p p^{max} \sum\limits_{i \in \mathcal{M}, i \neq m} \beta_{i,n}(t) |\boldsymbol \lambda_m^{H}(t) \boldsymbol \lambda_m(t)|^2 +1}.
\end{equation}
It can be easily shown that the minimum mean-square error channel estimate is unbiased and that the variance of the estimate $\boldsymbol{\hat{g}}^{a}_{mn}(t)$ is\cite{Guenach2021} 
\begin{equation}
    \vartheta_{mn}(t) = \frac{\tau_p p_m^p \beta^2_{mn}(t)}{\tau_p p_m^p \beta_{mn}(t) + \sigma^2_a}.
\end{equation}

\subsection{Data Transmission}

\subsubsection{Uplink Data Transmission}
In UCMEC, each user is served by a specific AP cluster. We define the AP cluster size for each user as $S$. The division of AP clusters is based on the estimated channel $\boldsymbol{\hat{g}_{mn}}(t)$ in descending order (i.e., selecting the top $\Xi$ APs with the best channel quality). For user $m$, the AP cluster at time slot $t$ is denoted as $\Phi_m(t)$. An AP can be part of the AP clusters of multiple users, thus serving multiple users simultaneously. Assume the set of users served by AP $n$ is denoted as $\Omega_n(t)$. The uplink payload signal received by AP $n$ at time slot $t$ is given by:
\begin{equation}
   \boldsymbol y_n(t) = \sum_{m \in \mathcal{M}} \sqrt{p_m^d(t)} \boldsymbol g^{a}_{mn}(t) x_m(t) + \boldsymbol z_n(t),
\end{equation}
where $p_m^d(t)$ represents the transmit power of user $m$ during uplink data transmission, and $x_m(t)$ denotes the quadrature amplitude modulation data symbols sent by user $m$. In this paper, we assume that the user's transmit power needs to be selected from a discrete set of power levels $\mathcal{P}^a = {0, 0.2, 0.4, ..., p^{max}}$ \cite{Huifa2015}. Users only transmit data to their own AP cluster, and therefore, only the APs within the cluster will decode the signal. Using maximum ratio combining (MRC), AP $n$ first partially equalizes the received signal based on the local channel estimates as $\boldsymbol{(\hat{g}}^{a}_{mn}(t))^H \boldsymbol y_n(t)$ and then sends it to the CPU for further decoding. Based on the signals sent by the APs, the CPU will sum up the different partially equalized signals for user $m$ as:
\begin{equation}
    \hat{y}_m(t) = \sum_{n \in \Phi_m(t)} \boldsymbol{(\hat{g}}^{a}_{mn}(t))^H \boldsymbol y_n(t),
\end{equation}
We can obtain the Signal-to-Interference-plus-Noise Ratio (SINR) of user $m$ at time slot $t$\cite{Guenach2021}: 
\begin{equation}\setlength{\abovedisplayskip}{3pt}  \setlength{\belowdisplayskip}{3pt}
    \zeta^{a}_m(t) = \frac{\rho p_m^d(t) (\sum\limits_{n \in \phi_m(t)} \vartheta_{mn}(t))^2}{ \sum\limits_{i \in \mathcal{M}, i \neq m}(\sum\limits_{n \in \phi_i(t)}\vartheta_{i,n}(t) \beta_{i,n}(t)) + \sum\limits_{n \in \Phi_m(t)} \sigma_a^2 \vartheta_{mn}(t)}.
\end{equation}
 In this paper, we assume that the user's transmit power needs to be selected from a discrete set of power levels $\mathcal{P}^a = \{0, 0.2, 0.4, ..., p^{max}\}$ \cite{Huifa2015}. Assume the bandwidth of the access channel is $B^{a}$, the maximum achievable rate for user $m$ at time slot $t$ can be calculated as $r^{a}_m(t) = B^{a}{\rm log_2}(1+\zeta^{a}_n(t))$.
\begin{equation}
    r^{a}_m(t) = B^{a}{\rm log_2}(1+\zeta^{a}_n(t)).
\end{equation}

\subsubsection{Fronthaul Data Transmission}
As mentioned above, for a specific user's offloading data, the CPU needs to recover the data based on the signals from all the APs in the AP cluster. Therefore, all the APs in each user's AP cluster need to transmit data to a specific CPU via wireless fronthaul. According to Eq. (8), the transmission rate when AP $n$ transmits data to CPU $k$ can be obtained using Shannon's formula:
\begin{equation}
    r^{f}_{nk}(t) = B^{f}{\rm log_2}(1+\zeta^{f}_{nk}(t)),
\end{equation}
where $B^{f}$ represents the fronthaul bandwidth.

\subsection{System Delay Analysis}
In this paper, we consider binary offloading, i.e., all the tasks cannot be divided. we use a binary variable $ \omega_{mk}(t)$ to indicate whether user $m$ needs to offload to CPU $k$ at time slot $t$, where $ \omega_{mk}(t)=1 $represents user $m$ chooses to offload to CPU $k$, otherwise it will be processed locally. Since the data size of the return results is negligible, we can ignore the downlink transmission process. For the decision variable $g_{nk}(t)$ in Sec. III. B, it is related to the decision variables $\omega_{mk}(t)$ and $\Phi_m(t)$ as follows: 
\begin{equation}
     g_{nk}(t)=\left\{
\begin{array}{rcl}
1      &      & \text{if} \quad\omega_{mk}(t)=1\quad \text{and AP $n$ $\in \Phi_m(t)$} \\
0     &      & \text{otherwise.}\\
\end{array} \right. 
\end{equation}

\subsubsection{Local Computing Delay}
The delay for user $m$ when processing the task locally at time slot $t$ can be obtained as follows:
\begin{equation}
    T_m^{l}(t) = \sum_{k \in \mathcal{K}} \left(1-\omega_{mk}(t) \right) \frac{D_m(t) \rho_m(t)}{C_m^{\rm l}}.
\end{equation}

\subsubsection{Task Offloading Delay}
When a user sends an offloading request to the CPU, the user first transmits the data signal to all the APs within the AP cluster through the access channel. Therefore, the transmission delay for user $m$ when offloading to the AP cluster $\Phi_m(t)$ can be given as: 
\begin{equation}
    T_m^{a}(t) = \frac{\omega_{mk}(t) D_m(t)}{r^{a}_m(t)}.
\end{equation}

Then, all the APs in the AP cluster will transmit the data of the served users to a specific CPU through the fronthaul channel. The data transmission delay for user $m$ in the fronthaul can be expressed as the maximum value of the fronthaul transmission delay of all the APs in the cluster, which can be given by:
\begin{equation}
    T_m^{f}(t) =  \sum_{k \in \mathcal{K}} \omega_{mk}(t)\max_{n \in \Phi_m(t)}\{ \frac{ D_m(t)}{r^{f}_{nk}(t)}\}.
\end{equation}

Assume the computing resource that CPU $k$ allocates to user $m$ is $c_{mk}(t)$, the task processing delay for user $m$ at the CPU can be expressed as follows:
\begin{equation}
    T_m^{e}(t) = \sum_{k \in \mathcal{K}} \omega_{mk}(t) \frac{ D_m(t)\rho_m(t)}{c_{mk}(t)}.
\end{equation}
Therefore, the offloading delay of user $m$ can be expressed as 
\begin{equation}
    T_m^{o}(t) = T_m^{a}(t) + T_m^{f}(t) + T_m^{e}(t).
\end{equation}
And we can obtain the total delay of user $m$:
\begin{equation}
    T_m(t) = \max\left\{T_m^l(t), T_m^o(t) \right\}.
\end{equation}

\subsection{Problem Formulation}
For efficient task offloading and resource allocation, users and CPUs need to incorporate the long term
impact of its actions into its decision. Therefore, we aim to minimize the long-term average total delay of all users. Assuming the offloading decision matrix, uplink transmit power control matrix and the computing resource allocation matrix are denoted by $\boldsymbol{\varOmega} = \left\{\omega_{mk}(t), m \in \mathcal{M}, k \in \mathcal{K}, t\in\mathcal{T}\right\}$, $\boldsymbol{p}= \left\{p_{m}^{d}(t), m \in \mathcal{M}, t\in\mathcal{T}\right\}$, $\boldsymbol C = \left\{c_{mk}(t), m \in \mathcal{M}, k \in \mathcal{K}, t\in\mathcal{T}\right\}$, respectively. We can formulate the following optimization problem:
\begin{subequations}\label{P1}
	\begin{align}
		\mathrm{P1}:
		&\quad \min_{\boldsymbol{\varOmega}, \boldsymbol{p}, \boldsymbol C} \frac{1}{MT}\sum_{t=1}^{T}\sum_{m=1}^{M}T_{m}(t),\\
		s.t.\;
            &\quad \sum_{k \in \mathcal{K}}\omega_{mk}(t) \leq 1, \forall m \in \mathcal{M}, \forall t\in \mathcal T \\
		&\quad \sum_{m \in \mathcal{M}}c_{mk}(t) = C^e_{k}, \forall k \in \mathcal{K}, \forall t\in \mathcal T, \\
		&\quad  p_m^{\rm d}(t) \in \mathcal{P}^a , \forall m\in \mathcal M, \forall t\in \mathcal T,  \\
		&\quad  |\Phi_m(t)| = \Xi, \forall m\in \mathcal M, \forall t\in \mathcal T, \\
          &\quad  T_m(t) \leq \tau_c, \forall m\in \mathcal M, \forall t\in \mathcal T, \\
		&\quad 0 \leq c_{mk}(t) \leq C^e_k, \forall m\in \mathcal M, k \in \mathcal{K}, \forall t\in \mathcal T, \\
		&\quad \omega_{mk}(t) \in、\{0,1\},\forall m\in \mathcal M, k \in \mathcal{K}, \forall t\in \mathcal T.
	\end{align}
\end{subequations}
where (24b) represents that users can offload their tasks to at most one CPU. (24c) ensures that the allocated computation resource will not exceed the computational capacity. (24d) specifies that the selected transmit power for each user should belong to the predefined power level set. (24e) imposes a limit on the size of each user's AP cluster. (24f) requires that all tasks of users must be completed within each time slot, otherwise will be discarded. (24g) and (24h) define the domain of the computing resource allocation variables and offloading decision variables, respectively. Upon observing the optimization problem, it can be noticed that it is a mixed-integer nonlinear programming (MINLP) problem with coupled variables. The optimization objective is non-convex, and this problem can be proven to be NP-Hard.\cite{Mengting2018}.

\section{Non-Cooperative Optimization Scheme}

Due to the limited resources in the network and the fact that users are typically self-interested, each user's decision is usually made to maximize their own benefits. In this section, we design a decentralized joint optimization framework under the assumption that users and CPUs will not cooperate with each other (in addition to necessary information transmission). Firstly, we can decompose problem (24) into two sub-problems: a continuous computation resource allocation sub-problem and a discrete task offloading and power control sub-problem. The continuous computing resource allocation sub-problem can be solved on the CPU side using convex optimization, while the task offloading and power level control sub-problem will be solved on the user side using multi-agent deep reinforcement learning.

\subsection{Computing Resource Allocation}
Given the user's offloading and power control decisions, the computing resource allocation decision at the CPU can be obtained by solving the following sub-problem:
\begin{subequations}\label{P2}
	\begin{align}
		\mathrm{P2}:
		&\quad \min_{\boldsymbol C} \frac{1}{MT}\sum_{t=1}^{T}\sum_{m=1}^{M}T_{m}(t)\\
		s.t.\;
		&\quad \sum_m^{\mathcal{M}} c_{mk}(t) = C^{e}_k, \forall k \in \mathcal{K}, \forall t\in \mathcal T, \\
            &\quad  T_m(t) \leq \tau_c, \forall m\in \mathcal M, \forall t\in \mathcal T, \\
		&\quad 0 \leq c_{mk}(t) \leq C_k^e, \forall k \in \mathcal{K}, \forall t\in \mathcal T. 
	\end{align}
\end{subequations}

We decouple the problem as a instantaneous problem in each time slot. Thus, in time slot $t$, the computing resource allocation decision at the CPU can be obtained by solving the following optimization problem:
\begin{subequations}\label{P3}
	\begin{align}
		\mathrm{P3}:
		&\quad \min_{\boldsymbol C(t)} \frac{1}{M}\sum_{m=1}^{M}T_{m}(t)\\
		s.t.\;
		&\quad \sum_m^{\mathcal{M}} c_{mk}(t) = C^{e}_k, \forall k \in \mathcal{K} \\
            &\quad  T_m(t) \leq \tau_c, \forall m\in \mathcal M,  \\
		&\quad 0 \leq c_{mk}(t) \leq C_k^e, \forall k \in \mathcal{K}. 
	\end{align}
\end{subequations}
Where $\boldsymbol C(t)={c_{mk}(t), m \in \mathcal{M}, k \in \mathcal{K}}$. By expanding the optimization objective and removing irrelevant variables, the above expression can be represented as the following optimization problem:
\begin{subequations}\label{P4}
	\begin{align}
		\mathrm{P4}:
		&\quad \min_{\boldsymbol C(t)} \sum\limits_{m=1}^{M}\max\left\{\sum\limits_{k \in \mathcal{K}}\left(1-\omega_{mk}(t)\right)\frac{D_m(t) \rho_m(t)}{C_m^l}, \nonumber\right.\\
  &\left.\max_{n \in \Phi_m(t)}\{\frac{D_m(t)}{r^{a}_m(t)}\} +  \max_{n\in \Phi_m(t)}\{\frac{D_m(t)}{r^{f}_{nk}(t)}\} + \frac{D_m(t) \rho_m(t)}{c_{mk}(t)}\right\} \\
		s.t.\;
		&\quad \sum\limits_m^{\mathcal{M}} c_{mk}(t) = C^{e}_k, \forall k \in \mathcal{K} \\
            &\quad  T_m(t) \leq \tau_c, \forall m\in \mathcal M, \\
		&\quad 0 \leq c_{mk}(t) \leq C_k^e, \forall k \in \mathcal{K}.
	\end{align}
\end{subequations}

 According to the properties of convexity preserving operations, it can be easily proved that the optimization objective and constraints with respect to $\boldsymbol C(t)$ are both convex, thus making the problem a convex optimization problem. Therefore, we can use convex optimization toolbox such as CVXPY \cite{cvxpy} to obtain the optimal solution $\boldsymbol C^{*}(t)$.

\subsection{Non-Cooperative Task Offloading and Power Allocation}

Next, we will introduce how to deploy MADRL algorithm on the user side to obtain task offloading and power allocation decisions. First, we model the user's decision-making process for task offloading and power allocation as a partially observable Markov decision process (POMDP). Then, based on the decision objectives of each agent, we propose an IPPO-based optimization scheme to obtain the optimal policy.

\subsubsection{Non-Cooperative POMDP Formulation}
We assume that each user will acted as an independent decision-making agent and treat other users as part of the environment. The task offloading and power control process of users can be modeled as a partially observable Markov decision process (POMDP) \cite{Zhan2020}, which comprises the following elements: $\left\{ \mathcal S, \mathcal A, r, \mathcal P, \mathcal{F}, \mathcal O, \gamma \right\}$. In each time slot $t$, $\mathcal S$ represents the global state set $s(t) \in \mathcal S$, which is unavailable for users. $\mathcal A = \left\{ \mathcal A_m, \forall m \in \mathcal{M} \right\}$ is the set of possible actions, and $\boldsymbol{a}_m(t)$ represents the action chosen by user $m$ in time slot $t$. $\mathcal P = \left\{ P(\boldsymbol s(t+1)| \boldsymbol s(t),\boldsymbol{a}(t)), \forall t \in \mathcal{T} \right\}$ is the set of state transition probabilities, describing the dynamic changes of the system. After making individual decisions based on their own objectives, each user has their own reward function $r_m(\boldsymbol s(t),\boldsymbol{a}(t)), \forall t \in \mathcal{T}$, which describes the reward obtained by user $m$ when taking a certain action in a specific state (abbreviated to $r_m(t)$ below). $\mathcal O = \left\{ \boldsymbol{o}_m(t), \forall m \in \mathcal{M}, \forall t \in \mathcal{T} \right\}$ represents the set of observations made by user $m$ in time slot $t$ regarding the global state $\boldsymbol s(t)$. $\mathcal{F} = {F_m: \boldsymbol{s}(t) \xrightarrow{} \boldsymbol{o}_m(t), \forall m \in \mathcal{M}, \forall t \in \mathcal{T} }$ is the set of observation functions. Each user $m$ generates individual observations $\boldsymbol{o}_m(t)$ based on the global state $\boldsymbol{s}(t)$. $\gamma \in [0,1]$ is the future discount factor, which weighs future rewards in the total reward. A larger $\gamma$ implies that the user values future rewards more, whereas a smaller $\gamma$ can be seen as more short-sighted.

At time $t$, user $m$ can only observe a partial state subset of $\boldsymbol s(t)$ and obtains its own observation set $\boldsymbol{o}_m(t)$. Based on this information, the user makes its decision action $\boldsymbol{a}_m(t)$. At this point, we can obtain the joint actions of all users $\boldsymbol{a}(t)$. Subsequently, the environment transitions to the next state $\boldsymbol{s}(t+1)$ based on the current state and joint actions, and provides a shared reward $r(t)$ to all users. This process is repeated continuously until the optimal decision is obtained, and the objective of each user is to maximize the total reward.

Below, we will provide the definitions of observation set, action set, and reward set for each agent in the non-cooperative task offloading and resource allocation scheme.

\begin{itemize}

\item{\textbf{Observation and State:} In this case, users cannot observe the decisions made by other users and can only treat other agents as part of the environment. Therefore, the decisions made by users may benefit or harm other agents. We define user $m$'s observation at time $t$ into three parts: individual demand information, historical decision information, and historical performance information. Individual demand information includes the task data size and computational demand in the current time slot. Historical decision information refers to the task offloading and power allocation decisions made by the user in the previous time slot. Historical performance information represents the total latency experienced by the user in the previous time slot. Thus, we define $\boldsymbol o_m(t) \triangleq \left\{D_m(t), \rho_m(t), \boldsymbol{a}_m(t-1), \forall t \in \mathcal{T} \right\}$.

We define the global state as the ensemble of observations of all users, i.e., $\boldsymbol{s}(t) = {\boldsymbol{o}_m(t), \forall m \in \mathcal{M}}$.
}

\item{\textbf{Action:} After obtaining observations, each user makes task partition and power control decisions for the current time slot based on the observation results. We define the action as two discrete variables, $\boldsymbol a_m(t) \triangleq \left[ \omega_{mk}(t), p_m^{d}(t), \forall m \in \mathcal{M}, \forall t \in \mathcal{T} \right]$.}

\item{\textbf{Reward:} In each time slot, after making decisions, each user receives a reward $r_m(t)$ based on the current state and the chosen actions. The reward for user $m$ at time $t$ is composed of two parts: the negative value of the total latency experienced by the user in the time slot, and a task processing latency penalty term. The penalty term is applied when the user's task is not completed within the current time slot, and is defined as $r_m(t) = -T_{m}(t) + \kappa^{c}(\tau_c - T_m(t))$, where $\kappa^c$ is the weight of the penalty term. Notably, the computation of the edge processing latency in the total latency requires the optimal computation resource allocation decision for the current time slot, which will be obtained through the algorithm described in Section 4.1.
}
\end{itemize}

\subsubsection{IPPO-based Algorithm Design}
To obtain the optimal policy of each user, we adopt independent learning-based IPPO\cite{IPPO} framework where each user apply proximal policy optimization (PPO) \cite{PPO} to learn decentralized policy. As a policy gradient-based DRL algorithm, PPO introduces the policy clipping mechanism to ensure training stability. Each PPO agent has two deep neural networks: an actor network taking the current observation state as input and generating actions as output to approximate the optimal policy, and a critic network evaluating the policy and approximating the value function. In PPO, the updated policy is used instead of the original policy to interact with the environment, ensuring that the samples can be reused. Additionally, PPO introduces the clipping mechanism that prunes the bias between the new policy and the original policy. When the bias between the new policy and the original policy exceeds a threshold, the policy gradient becomes zero, preventing parameter updates and ensuring algorithm stability.

The optimization framework based on IPPO is shown in Fig. 2. At the beginning of each time slot $t$, each user observes the task information and the history information. Next, the user uses the current observation as input to the actor network and obtains the current action, i.e., the power control and task offloading decision. After the task is processed and completed, The user then obtain its total delay (if the user selects task offloading, it will be returned from the CPU. Otherwise, it will be obtained by itself) and calculate the reward in this time slot, denoted as the reward $r(t)$, and stores the observation, action, and reward in the replay buffer. Each user updates its actor and critic network parameters every $B$ time slots. At this point, the user extracts $B$ samples as a mini-batch from its replay buffer and estimates the parameter gradients for the actor and critic networks. Subsequently, the user updates the network parameters based on the gradients, and the buffer is then cleared.

Next, we derive the policy update process of each user. For convenience, the slot index $t$ is omitted below. For user $m$, we assume that its policy is parameterized by $\boldsymbol \theta_m$, denoted as $\pi_m(\boldsymbol \theta_m)$. The policy optimization objective for user $m$ is to maximize the discounted return, which is equivalent to maximizing the expected value of the state-value function:
\begin{equation}
    \begin{aligned}
        \boldsymbol \theta^{*}_m &= \mathrm{argmax}_{\boldsymbol \theta_m} L_m(\pi_m(\theta_m)) \\
        &= \mathrm{argmax}_{\boldsymbol \theta_m} \mathbb{E} \left[ V_{\pi_m(\theta_m)}(\boldsymbol o_m(0)| \iota_m(0)) \right] \\
        &= \mathrm{argmax}_{\boldsymbol \theta_m} \mathbb{E} \left[ Q_{\pi_m(\theta_m)}(\boldsymbol o_m(0), \boldsymbol a_m(0)| \iota_m(0),\pi_m(\theta_m)) \right],
    \end{aligned}
\end{equation}
where $V_{\pi_m(\theta_m)}(\boldsymbol o_m)$ and $Q_{\pi_m(\theta_m)}(\boldsymbol o_m, \boldsymbol a_m)$ are the state-value function and the action-value function, respectively. And we have
\begin{equation}
    V_{\pi_m(\theta_m)}(\boldsymbol o_m) = \mathbb{E}\left[ R_m(t) | \boldsymbol o_m(t) = \boldsymbol o_m, \Pi, \mathcal{P}, \mathcal{F}   \right],
\end{equation}

\begin{equation}
    Q_{\pi_m(\theta_m)}(\boldsymbol o_m, \boldsymbol a_m) = \mathbb{E}\left[ R_m(t) | \boldsymbol o_m(t) = \boldsymbol o_m, \boldsymbol a_m(t)= \boldsymbol a_m, \Pi, \mathcal{P}, \mathcal{F}   \right],
\end{equation}

\begin{equation}
   R_m(t) = \sum_{j=t}^{T} \gamma(j-t)r_m(t).
\end{equation}

In the above equation, $\Pi = \left\{ \pi_m(\boldsymbol \theta_m), \forall m \in \mathcal{M} \right\}$ represents the policy set for all users. $V_{\pi_m(\theta_m)}$ is the value function observed by user $m$. $Q_{\pi_m(\theta_m)}$ represents the value function based on observations and actions. $\iota_m(0)$ is the initial observation probability distribution for user $m$. $R_m(t)$ is the discounted expected future reward for user $m$ at time slot $t$. The policy gradient\cite{Li2021} for each user can be obtained as follows:
\begin{equation}
\begin{aligned}
     &\nabla_{\boldsymbol \theta_m}L_m = \mathbb{E}_{\pi_m(\boldsymbol \theta_m), \iota_m(1)}\left[     \nabla_{\boldsymbol \theta_m} \rm{log} \pi_m(\boldsymbol \theta_m) Q_{\pi_m(\theta_m)}(\boldsymbol o_m, \boldsymbol a_m)    \right] \\
     &=  \mathbb{E}_{\pi_m(\boldsymbol \theta_m), \iota_m(1)}\left[     \nabla_{\boldsymbol \theta_m} \rm{log} \pi_m(\boldsymbol \theta_m) A_{\pi_m(\theta_m)}(\boldsymbol o_m, \boldsymbol a_m)    \right] \\
     &\approx  \mathbb{E}_{\pi_m(\boldsymbol{\hat{\theta}_m}), \iota_m(1)}\left[ f_m    \nabla_{\boldsymbol \theta_m} \rm{log} \pi_m(\boldsymbol \theta_m) A_{\pi_m(\theta_m)}(\boldsymbol o_m, \boldsymbol a_m)    \right],
\end{aligned}
\end{equation}
where $f_m = \frac{\pi_m(\boldsymbol \theta_m)}{\pi_m(\boldsymbol{ \hat{\theta}_m)}}$, $ A_{\pi_m(\theta_m)}(\boldsymbol o_m, \boldsymbol a_m) =  Q_{\pi_m(\theta_m)}(\boldsymbol o_m, \boldsymbol a_m) -  V_{\pi_m(\theta_m)}(\boldsymbol o_m, \boldsymbol a_m)$ is the advantage function for observation and action, which is estimated by the generalized advantage estimation (GAE)\cite{Schulman2015}, $ \boldsymbol{\hat{\theta}_m}$is the policy parameter by sampling, $\iota_m(1)$ is the observation distribution. According to \cite{John2017}, we can clip policy gradient as：
\begin{equation}
    \nabla_{\boldsymbol \theta_m}L_m \approx \mathbb{E}_{\pi_m(\boldsymbol{ \hat{\theta}_m}), \iota_m(1)}\left[ \nabla_{\boldsymbol \theta_m} \rm{log} \pi_m(\boldsymbol \theta_m) C(\boldsymbol o_m, \boldsymbol a_m)    \right],
\end{equation}
where，
\begin{equation}
\begin{aligned}
    C(\boldsymbol o_m, \boldsymbol a_m) = \min [&   f_m A_{\pi_m(\theta_m)}(\boldsymbol o_m, \boldsymbol a_m), \upsilon(f_m) A_{\pi_m(\theta_m)}\\&(\boldsymbol o_m, \boldsymbol a_m)    ],
\end{aligned}
\end{equation}

\begin{figure*}[tbp]
	\centering
	\includegraphics[angle=0,scale=0.25]{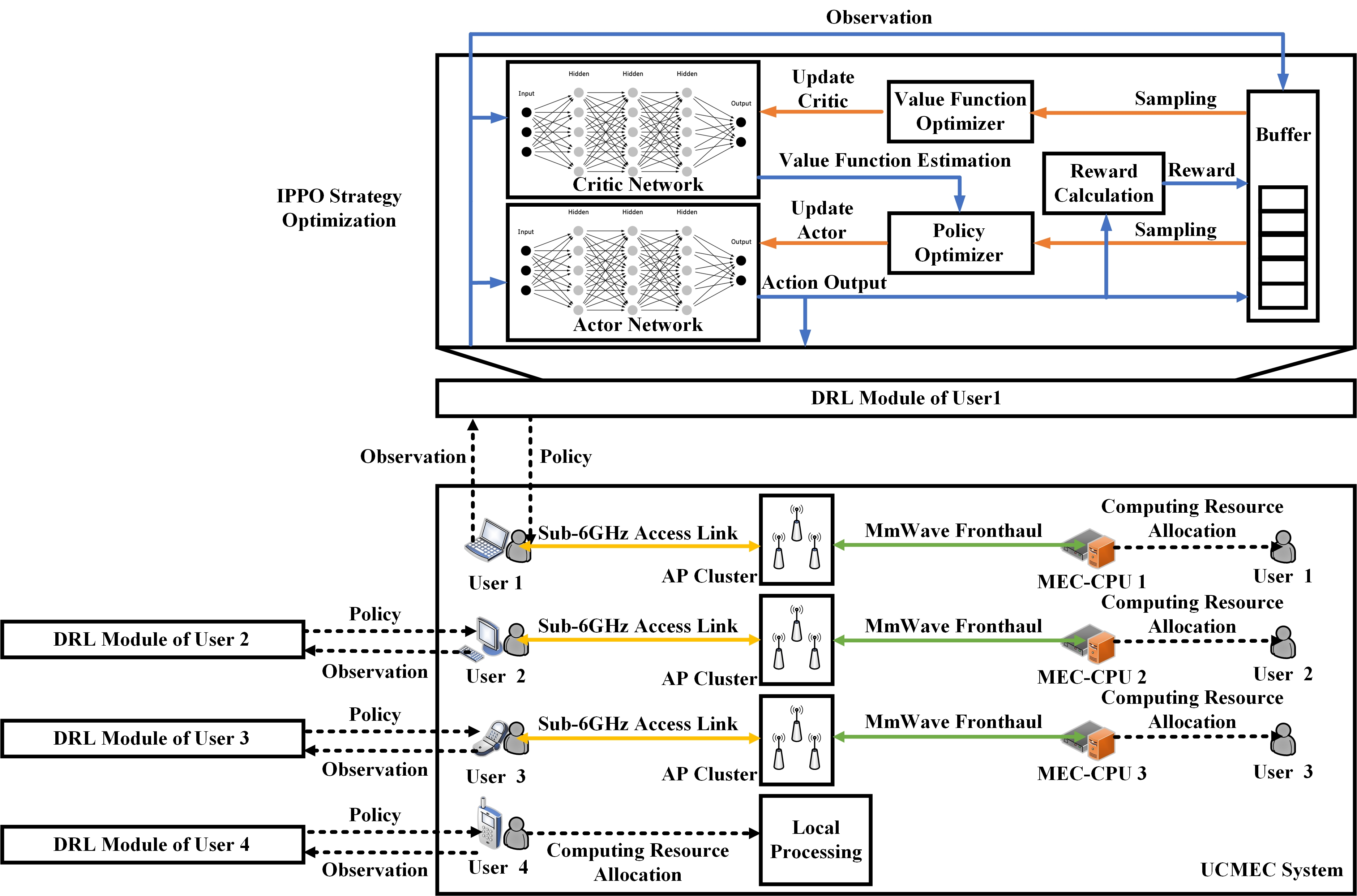}
	\caption{Non-Cooperative Optimization Framework}\label{fig:application}
\end{figure*}

\begin{equation}
 \upsilon(x)=\left\{
\begin{array}{rcl}
&1+\epsilon,  &x>1+\epsilon,\\
&x,           & 1-\epsilon \leq x \leq 1+\epsilon,\\
& 1-\epsilon, & x<1-\epsilon.
\end{array} \right. 
\end{equation}
where $\epsilon$ is an adjustable parameter. We define the loss function for updating the critic network of user $m$ as follows:
\begin{equation}\setlength{\abovedisplayskip}{3pt}  \setlength{\belowdisplayskip}{3pt}
\begin{aligned}
        J_m(\boldsymbol \phi_m) &= \mathbb{E}_{\boldsymbol o_m \sim \iota_m(1) }[  -V_{\boldsymbol \phi_m}(\boldsymbol o_m) \\
        &+ \mathbb{E}_{\boldsymbol o_m^{'}, \boldsymbol a_m}[   r + V_{\boldsymbol \phi_m}(\boldsymbol o_m^{'}) ] ]^2,
\end{aligned}
\end{equation}
where $\boldsymbol \phi_m(t)$ represents the parameters of the critic network, and $\boldsymbol o_m^{'}$ is the observation result of user $m$ at the next time step. The estimated gradient for the critic network parameters is calculated as:
\begin{equation}
    \nabla_{\boldsymbol \Phi_m(t)}J^{l}_m = \frac{1}{B} \sum_{t=0}^{B-1}\left[ V_{\boldsymbol \Phi_m(t)(\boldsymbol o_m(t)) - Y_m(t)} \frac{\mathrm{d}V_{\boldsymbol \Phi_m(t)}(\boldsymbol o_m)  }{\mathrm{d} \boldsymbol \Phi_m(t)}    \right],
\end{equation}
where
\begin{equation}
    Y_m(t) = R_m(t) - \gamma^{B-t}R_m(B) + \gamma^{B-t}V_{\boldsymbol \Phi_m(t)}(\boldsymbol o_m(B)),
\end{equation}
where $B$ is the mini-batch size used in updating the critic network. Therefore, the parameters of the critic network are updated using mini-batch stochastic gradient descent:
\begin{equation}
    \boldsymbol \Phi_m(t) \gets \Phi_m(t) - l_m^{c} \nabla_{\boldsymbol \Phi_m(t)} J^{'}_m,
\end{equation}
Where $l_m^{c}$ is the learning rate for user $m$ to update the critic network, and $\boldsymbol \theta_m$ represents the parameter of the actor network. Similarly, the estimated gradient for the actor network is calculated as:
\begin{equation}
    \nabla_{\boldsymbol \theta_m}L_m^{'} = \frac{1}{B}\sum_{t=0}^{B-1}\nabla_{\boldsymbol \theta_m}\rm{log}\pi_m(\boldsymbol \theta_m)C(\boldsymbol o_m(t), \boldsymbol a_m(t)),
\end{equation}
Therefore, the update process for the parameter $\boldsymbol \theta_m$ using mini-batch stochastic gradient descent is as follows:
\begin{equation}
    \boldsymbol \theta_m \gets \boldsymbol \theta_m + l_m^{a}\nabla_{\boldsymbol \theta_m}L_m^{'},
\end{equation}
In summary, the overall offline training process of IPPO-based task offloading and resource allocation scheme is shown in Algorithm 1. In the offline training phase, by constructing a simulated UCMEC network environment, agents can obtain the optimal task offloading and resource allocation strategies. During the online deployment phase, agents can be directly deployed for online decision-making, or for online decision-making and fine-tuning training.

\begin{algorithm}[htbp]
\SetKwData{Left}{left}\SetKwData{This}{this}\SetKwData{Up}{up} \SetKwFunction{Union}{Union}\SetKwFunction{FindCompress}{FindCompress} \SetKwInOut{Input}{input}\SetKwInOut{Output}{output}
	 \BlankLine
	 {\textbf{Initialize} the UCMEC system environment and training parameters}\;
   \For{time slot $t \in \mathcal{T}$}{
	 \For{each user $m \in \mathcal{M}$}{
	Observe the UCMEC environment and update observation $\boldsymbol o_m(t)$\; 	
        Obtain the task offloading and power control decisions $\boldsymbol a_m(t) = \{p_m^d(t), \omega_{mk}(t), \forall k \in \mathcal{K}\}$\; }
    Calculate the current reward $r_m(t)$\;
    Store $\left\{ \boldsymbol o_m(t), \boldsymbol a_m(t), r_m(t), \boldsymbol o_m(t+1) \right\}$ in the replay buffer $\mathcal{B}_m$\;
    \If{$t \% B==0$}
    {Compute the estimated gradient of the actor network and the critic network  $\nabla_{\boldsymbol \theta_m}L_m$  and $\nabla_{\boldsymbol \theta_m}J_m$ by Eq. (37) and Eq. (40)\;
    Update the actor network $\boldsymbol \theta_m$ and the critic network $\boldsymbol \Phi_m(t)$  by Eq. (39) and Eq. (41)\;
   }}
 \caption{IPPO-based Task Offloading and Power Control Scheme}
\label{algo_disjdecomp}
\end{algorithm}

\section{Cooperative Optimization Scheme}

In some case, the users, APs and CPUs in the system have high mutual trust and a low probability of malicious behavior. At this point, network nodes are willing to collaborate with each other to increase the overall total benefit. In this section, we propose a decentralized joint optimization scheme based on node cooperation. Specifically, information sharing is allowed among users, and user can obtain historical information from other users as a reference during decision-making. The strategy of task offloading and power control will be obtained through MAPPO-based MADRL algorithm. In addition to solve computing resource allocation problem by convex optimization, the CPU is responsible for actor and critic training to make better use of the global information.

\subsection{Cooperative Task Offloading and Power Allocation}

\subsubsection{Cooperative POMDP Formulation}
In some case, network nodes are willing to collaborate with each other to increase the overall total benefit because of the high mutual trust and the low probability of malicious behavior\cite{Qimei2023}. In this section, we propose a decentralized joint optimization scheme based on node cooperation. Specifically, information sharing is allowed among users as a reference during decision-making. In addition to solve computing resource allocation problem by convex optimization, the CPU is responsible for gathering the state information and training actor and critic networks to make better use of the global information. Similar to Sec. IV, we can model the task offloading and power control process as a cooperative POMDP, and the observation, state, action, and the reward are defined as below.

\begin{itemize}
\item{\textbf{Observation and State}
  In this case, on top of the local observation histories, agents are able to share each other’s policies and experiences during decision-making process. Let $\boldsymbol{a}_{-m}(t)$ represent the actions of all users except user $m$ at time slot $t$. Assume the observation of user $m$ is $\boldsymbol o_m(t) \triangleq \left\{D_m(t), \rho_m(t), \boldsymbol a_m(t-1),\boldsymbol{a}_{-m}(t-1),T(t-1)\right\}$, where $T(t-1) = \frac{1}{M}\sum\limits_m^M T_m(t-1)$. We define the global state as the ensemble of observations of all users, i.e., $\boldsymbol{s}(t) = \{\boldsymbol{o}_m(t), \forall m \in \mathcal{M}\}$.}

\item{\textbf{Action:}
Similar to the non-cooperative case, after obtaining the observation, each user makes task offloading and power control decisions for the current time slot. We define the action as two discrete variables, i.e., task offloading and power control variables: $\boldsymbol a_m(t) \triangleq \left[ \omega_{mk}(t), p_m^{d}(t), \forall m \in \mathcal{M}\right]$}。

\item{\textbf{Reward:}
The reward $r(t)$ shared by all users at time $t$ consists of two components: the negative value of the average total delay of all users in the current time slot, and a penalty term for the task processing delay of all users. Thus, we have $r(t) = -\frac{1}{M}\sum\limits_{m}^{M}T_{m}(t) + \kappa^{n}\frac{1}{M}\sum\limits_{m}^{M}(\tau_c - T_m(t))$, where $\kappa^n$ is the weight of the penalty term.}
\end{itemize}


\subsubsection{MAPPO-based Algorithm Design}
Due to the limited ability of each agent to make decisions based on its own observations and treating other agents as part of the environment, it is inevitable to ignore the effective information of the overall environment, leading to low training efficiency and difficulty in ensuring the effectiveness of the training results. Firstly, from the perspective of a given agent, the presence of other learning and exploring agents makes the resulting environment unstable, leading to the loss of convergence guarantee. Secondly, independent learners may not always distinguish between the randomness of the environment and the exploration of another agent, which makes them unable to learn the optimal policy in certain environments. In fact, decentralized policies do not necessarily need to be learned in a decentralized way. For efficiency reasons, MARL training can be centralized, allowing agents to access each other's observations and other unobservable additional state information during training. Centralized training allows for joint training of all agents with a single joint policy conditioned on joint observations and additional state information. Therefore, in scenarios where information sharing among users is allowed, we adopt the centralized training and decentralized execution (CTDE) method of PPO, i.e., MAPPO\cite{MAPPO} to obtain the policies of users. In this case, each user only maintaining an actor network, and a powerful CPU can be selected as the central critic. Alternatively, multiple CPUs can be used as a central critic, and training can be conducted using distributed training methods such as federated learning.

When users are willing to collaborate with each other, all users have a common optimization objective, which is to determine the optimal distributed policy to maximize the discounted return. In MAPPO, the global state is available at the centralized critics for training. The centralized critic estimates the joint value function based on global information. Each user makes decisions only based on its local observation. After the training process, the global information is no longer needed, and users can choose actions in a decentralized manner. Assume the parameters of the actor network be $\boldsymbol \theta$ and the parameters of the critic network be $\phi$. At time slot $t$, the actor network maps the observation $\boldsymbol \omega_m(t)$ of user $m$ according to the current policy $\pi$ to a categorical distribution on the discrete action space, or to the mean and standard deviation vectors of a Multivariate Gaussian Distribution, from which an action is sampled \cite{MAPPO}. 

\begin{figure*}[tbp]
	\centering
	\includegraphics[angle=0,scale=0.25]{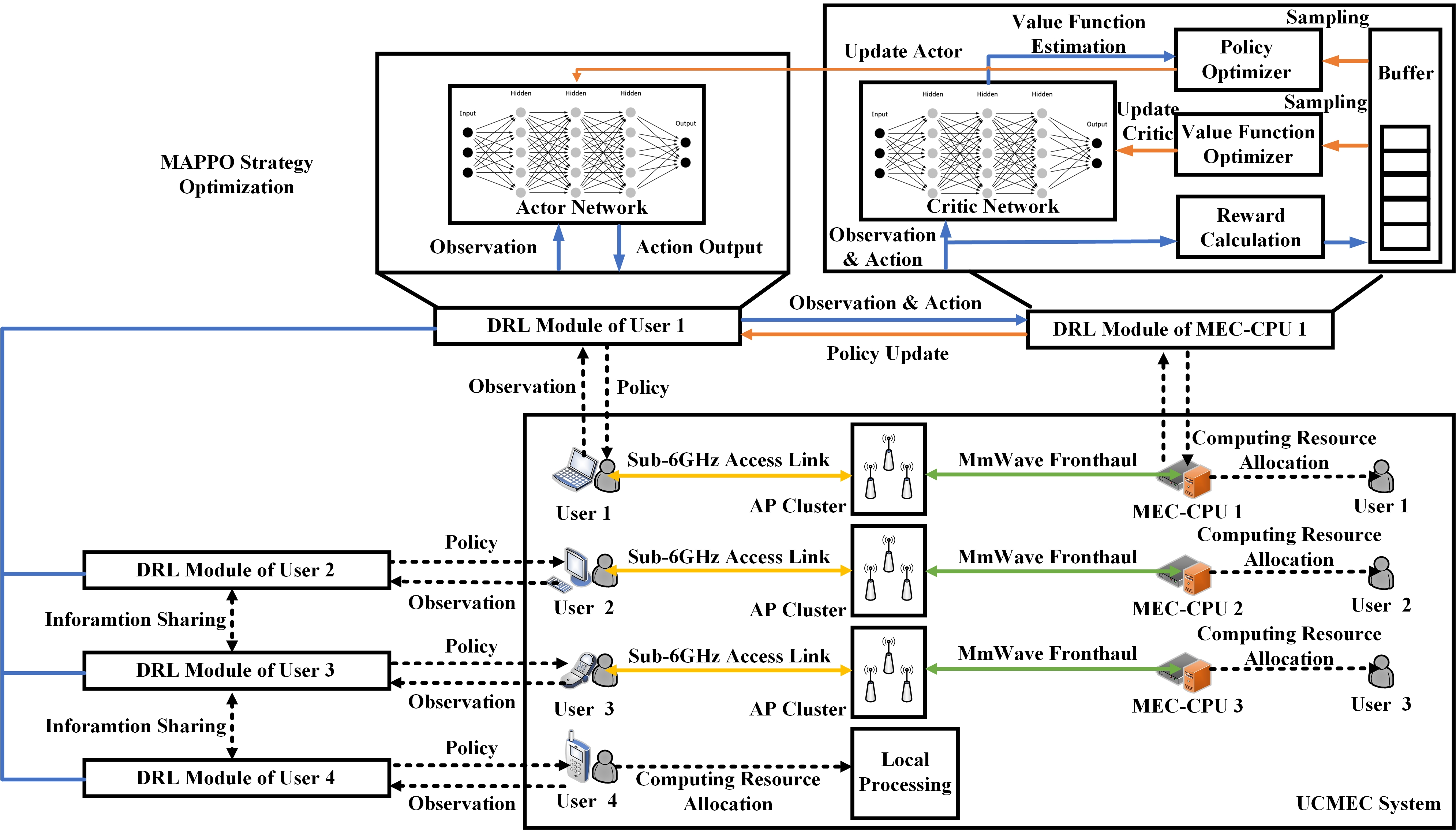}
	\caption{Cooperative Optimization Framework}\label{fig:application}
\end{figure*}

In MAPPO, we define the loss function for the actor network as follows:
\begin{equation}
\begin{aligned}
    &L(\boldsymbol \theta) = \frac{1}{BM} \sum\limits_{i=0}^{B-1} \sum\limits_{m=1}^{M}\min[ f_m A_{\pi_m(\boldsymbol \theta_m)}(\boldsymbol s(i),\boldsymbol a(i)), \\
    &\upsilon(f_m) A_{\pi_m(\boldsymbol \theta_m)}(\boldsymbol s(i),\boldsymbol a(i))]
     + \varrho\frac{1}{BM}\sum\limits_{i=0}^{B-1} \sum\limits_{m=1}^{M} S_{\pi_m(\boldsymbol \theta_m)}(\boldsymbol s(i)),
\end{aligned}
\end{equation}

where S is the policy entropy,  $\varrho$ is the entropy coefficient hyper-parameter. Then we can obtain the loss function of the critic network:
\begin{equation}
\begin{aligned}
        J(\boldsymbol \phi) &= \frac{1}{BM}\sum\limits_{i=0}^{B-1} \sum\limits_{m=1}^{M}\max[(V_{\boldsymbol \phi} (\boldsymbol s(i)) - \hat {R}(i))^2,\\
        &(\upsilon^{'}(V_{\boldsymbol \phi}(\boldsymbol s(i)))- \hat {R}(i)))^2],
\end{aligned}
\end{equation}
 and we have
\begin{equation}
\begin{aligned}
     &\upsilon^{'}(V_{\boldsymbol \phi}(\boldsymbol s(i)))\\
     &=\left\{
\begin{array}{rcl}
&V_{\boldsymbol \phi_{old}}(\boldsymbol s(i))+\epsilon,  &V_{\boldsymbol \phi}(\boldsymbol s(i)) > V_{\boldsymbol \phi_{old}}(\boldsymbol s(i)) +\epsilon,\\
&V_{\boldsymbol \phi}(\boldsymbol s(i)),           & V_{\boldsymbol \phi_{old}}(\boldsymbol s(i))-\epsilon \leq V_{\boldsymbol \phi}(\boldsymbol s(i)) \\
& &\leq V_{\boldsymbol \phi_{old}}(\boldsymbol s(i))+\epsilon,\\
& V_{\boldsymbol \phi_{old}}(\boldsymbol s(i))-\epsilon, & V_{\boldsymbol \phi}(\boldsymbol s(i)) < V_{\boldsymbol \phi_{old}}(\boldsymbol s(i))-\epsilon,
\end{array} \right.
\end{aligned}
\end{equation}
where $\hat {R}(i)$ is the discounted reward-to-go, $\phi_{old}$ are old parameters before the update. Through the loss function, we can calculate the gradient of actor and critical network, and update the parameters of actor and critical network similar to the gradient descent methods as Eq. (39) and Eq. (41).

The optimization scheme based on MAPPO is shown in Fig. 3. At the beginning of each time slot $t$, each user will first observe the environment, including task information, self historical information, and historical information of other users. Next, the user uses the current observation as input to the actor network and obtains the current task offloading and power control decisions. The user then sends the action and the current state to the CPU. The CPU integrates all users' information for computing resource allocation and obtains the current global state and actions for calculating the reward. Subsequently, the CPU stores the observation, action, and reward in the replay buffer. The CPU updates its actor and critic network parameters every $B$ time slots and informs the users about the updates in the actor network. Then, each user updates its own actor network parameters. Next, we will explain how the CPU updates the parameters of the actor and critic based on the global information. In summary, the overall offline training process of MAPPO-based task offloading and resource allocation scheme is shown in Algorithm 2.

\begin{algorithm}[htbp]
\SetKwData{Left}{left}\SetKwData{This}{this}\SetKwData{Up}{up} \SetKwFunction{Union}{Union}\SetKwFunction{FindCompress}{FindCompress} \SetKwInOut{Input}{input}\SetKwInOut{Output}{output}
	 \BlankLine
	 {\textbf{Initialize} the UCMEC system environment and training parameters}\;
   \For{time slot $t \in \mathcal{T}$}{
	 \For{each user $m \in \mathcal{M}$}{
	Observe the UCMEC environment and other users, then update observation $\boldsymbol o_m(t)$\; 	
        Obtain the task offloading and power control decisions $\boldsymbol a_m(t) = \{p_m^d(t), \omega_{mk}(t), \forall k \in \mathcal{K}\}$\;
          \If{$t \% B==0$}
    {
    Update the actor network $\boldsymbol \theta_m$  by Eq. (20)\;
    Save the network model\;
    }
        }
	 	\For{each CPU $k \in \mathcal{K}$}{
    Calculate the current reward $r(t)$\;
    Store $\left\{ \boldsymbol s(t), \boldsymbol a(t), r(t), \boldsymbol s(t+1) \right\}$ in the replay buffer $\mathcal{B}_m$\;
    \If{$t \% B==0$}
    {Compute the estimated gradient of the actor network and the critic network  $\nabla_{\boldsymbol \theta_m}L_m$  and $\nabla_{\boldsymbol \theta_m}J_m$ according to Eq. (42) and Eq. (43)\;
    Update and the critic network $\boldsymbol \Phi_m(t)$  according to Eq. (39) and Eq. (41)\;
     Send the actor network gradient to users. Clear the replay buffer $\mathcal{B}$ and save the network model\;
    }
     return the results to users after finishing task processing\;
    }
   }
                \caption{MAPPO-based Task Offloading and Power Control Scheme}
 	 	  \label{algo_disjdecomp}
 \end{algorithm}

\section{Performance Evaluation}

\subsection{Simulation Settings}


In this section, extensive simulations are conducted to validate the proposed optimization schemes in UCMEC. We implement the simulations using Python 3.10 and Pytorch 1.12 on a computer with one GeForce RTX 3080 GPU and an Intel Core i7-10700 CPU running on a processor speed of $2.9$ GHz, and $32$ GB RAM. In UCMEC, we assume there are 3 CPUs placed in (300m,300m), (600m,300m),(450m,600m). All APs and users are randomly distributed in a square area of $900 \times 900$ $\rm{m^2}$. The computing resource of the edge server deployed on the CPU are evenly distributed in $[10,20]$ CPU cycles frequency in GHz, and the local computing resource of each user are evenly distributed in $[2,5]$ CPU cycles frequency in GHz. We assume that the maximum transmit power of each user is $100$ mW. The user's computing task data size in each time slot is evenly distributed in $[50,100]$ KB, the computing density is evenly distributed in $[500,1000]$ CPU cycles/bit. For the access  channel model, we assume that the coherent time $\tau_c$ is $100 ms$, the noise power is $-174$ dBm/Hz, the carrier frequency is $1.9$ GHz, the bandwidth is $20$ MHz, the antenna heights of APs and users are $15$ m and $1.65$ m, respectively. We assume that the standard deviation of shadow fading is $8$ dB, the parameter of shadow fading $\delta$ is 0.5 the distance thresholds of the three-segments path-loss model are $10$ m and $15$ m. For the MmWave fronthaul channel, we assume the directivity gain of main lobes and side lobes are $10$ dB and $-10$ dB, respectively. The block density is $6\times10^{-4}$, the main lobe beamwidth is $30^{\circ} $ and the path-loss exponent of LoS and NLoS links are 2.5 and 4, respectively. We assume the Nakagami fading parameter of LoS and NLoS links are 3 and 2, respectively.

For each episode, we assume the episode length $T$ is $300$. For IPPO and MAPPO, both the actor and critic network have two fully connected hidden layers, each hidden layer contains $64$ neurons. The parameters of the neural networks are initialized orthogonally, and the generalized advanced estimator (GAE) method is adopted to calculate the policy gradient.  The weight of the penalty terms $\kappa ^c$ and $\kappa ^n$ in the reward function are set to $0.6$ and $0.8$, respectively. In addition, the learning rate of actor and critic networks is initialized to $5\times10^{-4}$. We set the discount factor to $0.99$, the GAE parameter to $0.95$, and the PPO clip parameter to $0.2$. We set the replay buffer size and the batch size to $2048$ and $64$, respectively. The simulation code is available on \href{https://github.com/qlt315/UCMEC-mmWave-Fronthaul}{https://github.com/qlt315/UCMEC-mmWave-Fronthaul}.


\begin{figure*}[htbp]
    \centering
    \captionsetup{justification=centering}
	  \subfloat[Convergence performance.]{
       \includegraphics[width=0.5\linewidth]{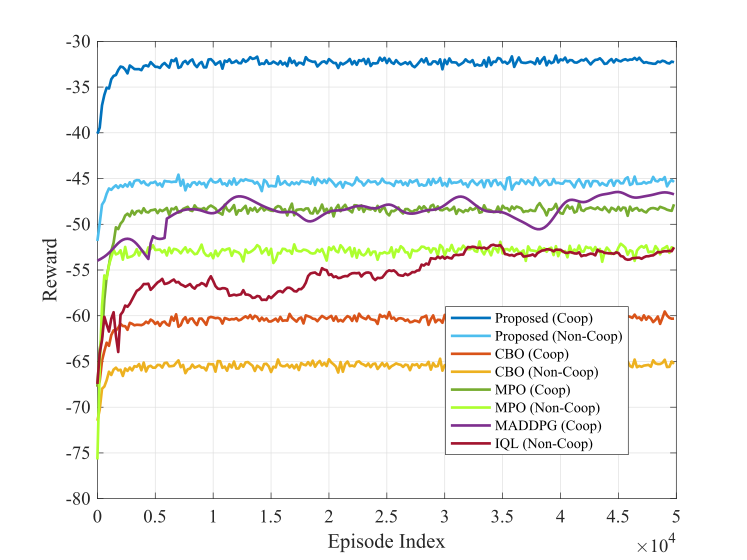}}\hspace{-8mm}
    \label{1a}
	  \subfloat[offline training time.]{
        \includegraphics[width=0.5\linewidth]{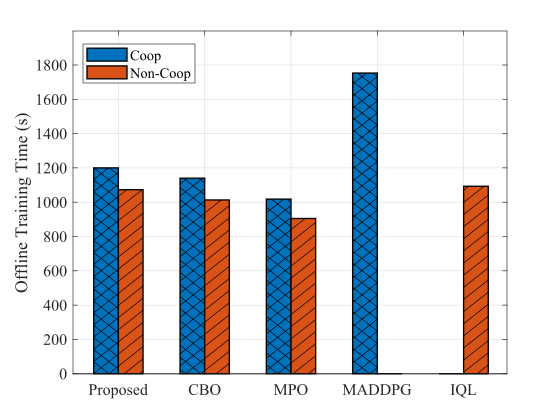}}
    \label{1b}
    	 \caption{Convergence performance and offline training time comparison.}
	\end{figure*}
\subsection{Comparison Schemes}
To verify the effectiveness of our proposed joint optimization scheme, we consider the following four reference schemes:

\begin{itemize}
\item{Cellular-Based Offloading (CBO)\cite{Ngo2017}}: To verify the effectiveness of the user-centric transmission method, we consider a small-cell system in this benchmark. The user will select an AP for uplink data transmission. It is assumed that each user  will choose the AP with the best channel conditions based on the channel estimation results. For fairness, the overhead of the fronthaul transmission can be ignored.

\item{Multi-Agent Deep Deterministic Policy Gradient (MADDPG)\cite{Ke2022,Qimei2023}}: MADDPG extends the deep deterministic policy gradient (DDPG) to multi-agent environments by introducing a centralized training with decentralized execution framework. Gumbel-Softmax sampling is applied to handle discrete actions of agents\cite{Lowe2017}.

\item{Independent Q-Learning (IQL)\cite{Zhao2022}}: IQL extends deep Q-learning (DQN) to multi-agent environments, where each agent runs the DQN algorithm\cite{IQL}. DQN is a model-free DRL algorithm. By using neural network to represent the traditional Q-table, DQN can ensure finding an optimal strategy in any finite Markov decision process (MDP).

\item{Maximum Power Offloading (MPO)\cite{Zilong2020}: To verify the necessity of optimizing transmission power allocation decisions in UCEC, all users use the maximum power to transmit the offloading task.}

\end{itemize}

\subsubsection{Convergence and Offline Training Time}

\subsection{Simulation Results}

\begin{figure*}[tbp]
    \centering
    \captionsetup{justification=centering}
	  \subfloat[Uplink transmission rate versus number of users.]{
       \includegraphics[width=0.5\linewidth]{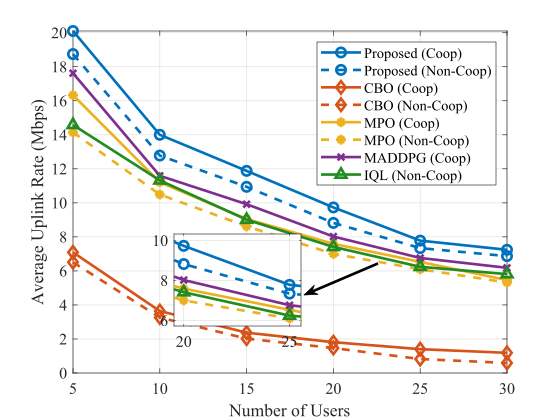}}\hspace{-8mm}
    \label{1a}
	  \subfloat[Uplink transmission rate versus number of APs.]{
        \includegraphics[width=0.5\linewidth]{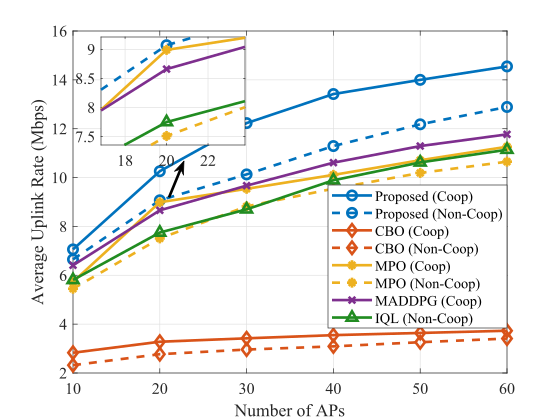}}
    \label{1b}
	  \subfloat[Uplink transmission rate versus maximum transmit power.]{
        \includegraphics[width=0.5\linewidth]{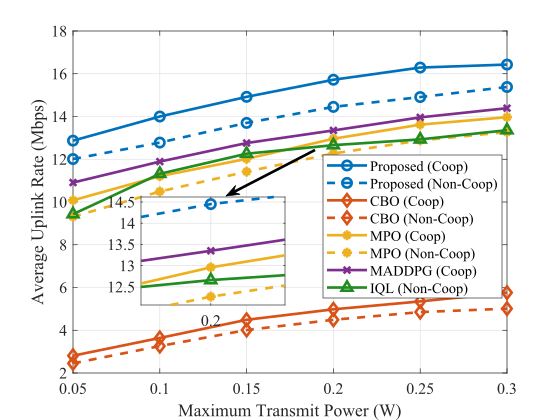}}\hspace{-8mm}
    \label{1c}
     \subfloat[Uplink transmission rate versus AP cluster size.]{
       \includegraphics[width=0.5\linewidth]{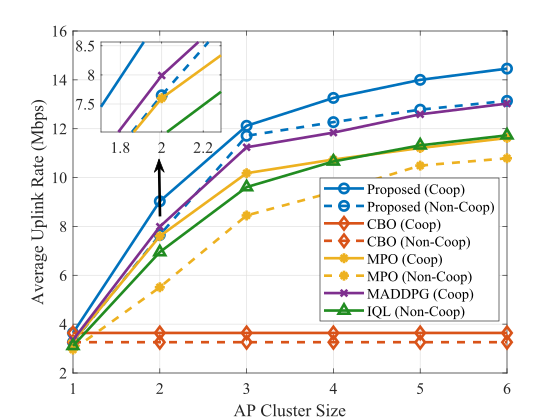}}
    \label{1a}
	 \caption{Uplink transmission rate under different parameters.}
	  \label{fig1} 
	\end{figure*}
Fig. 4(a) shows the convergence performance of different schemes, with 10 users and 50 APs. The figure reveals that schemes based on MAPPO and IPPO, namely Proposed, CBO, and MPO, can converge in approximately 3000 episodes, with a relatively stable convergence curve. This stability is achieved through the policy clipping mechanism of PPO, which ensures stability both before and after policy updates. In contrast, the convergence curves of IQL and MADDPG exhibit more oscillatory behavior, particularly for MADDPG, which faces challenges in achieving stable convergence due to its complex parameter adjustment. Surprisingly, for IL-based IPPO, its convergence performance is nearly on par with MAPPO. Additionally, the reward of non-cooperative schemes is lower than that of cooperative schemes because agents cannot benefit from global information and are more susceptible to falling into local optima. Fig. 4(b) shows the offline training time of 50000 steps for different schemes. The figure indicates that the training time of IL-based schemes in the non-cooperative case is shorter than that of CTDE-based schemes in the cooperative case. This difference arises because the centralized Critic architecture in the CTDE architecture requires global information for training, while the IL architecture makes decisions and undergoes training in a completely distributed manner. Furthermore, MADDPG exhibits a long training time due to its numerous hyper-parameters, while IQL has a slightly longer training time compared to IPPO. Moreover, CBO and MPO do not require AP clustering or power control decisions, resulting in a slightly shorter training time compared to the Proposed scheme. However, this reduction in training time comes at the cost of sacrificing system performance.

\subsubsection{Uplink Transmission Rate Performance}

\begin{figure*}[tbp]
    \centering
    \captionsetup{justification=centering}
	  \subfloat[Average total delay versus number of users.]{
       \includegraphics[width=0.5\linewidth]{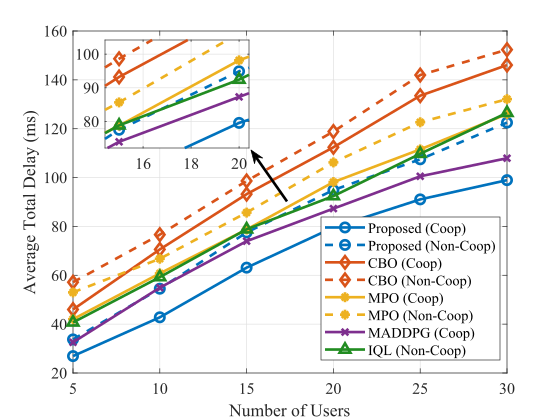}}\hspace{-8mm}
    \label{1a}
	  \subfloat[Average total delay versus number of APs.]{
        \includegraphics[width=0.5\linewidth]{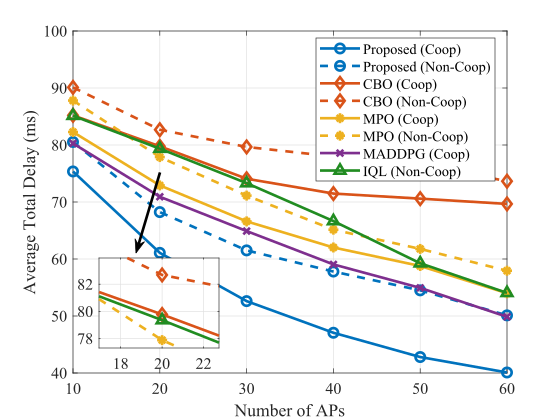}}
    \label{1b}
	  \subfloat[Average total delay versus maximum transmit power.]{
        \includegraphics[width=0.5\linewidth]{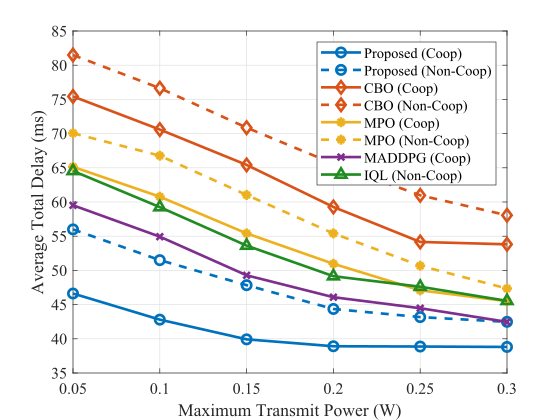}}\hspace{-8mm}
    \label{1c}
     \subfloat[Average total delay versus AP cluster size.]{
       \includegraphics[width=0.5\linewidth]{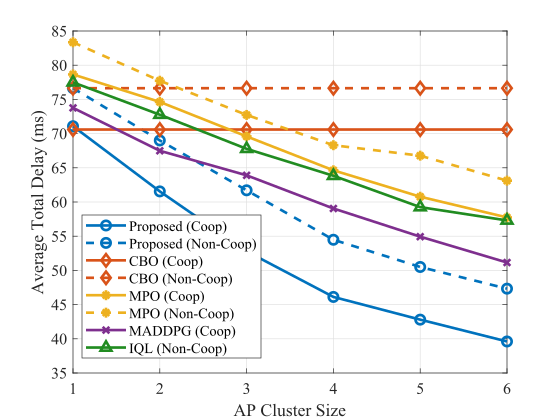}}
    \label{1a}
	 \caption{Average total delay under different parameters.}
	  \label{fig1} 
	\end{figure*}

Fig. 5 illustrates the comparison of the average uplink transmission rate of users under different parameters. In Fig. 5(a), with the number of APs fixed at 50, the average uplink rate of users decreases as the number of users increases. This reduction is attributed to the increased offloading requests from users, leading to greater interference during uplink transmission. The proposed algorithm achieves the highest transmission rate, followed by MADDPG, MPO, IQL, and CBO. Specifically, Proposed (Coop) at most exhibits an increase in transmission rate of 7.31\% - 208.76\% compared to other schemes. This superiority is due to AP cluster collaboration, providing services more resilient to interference and fading than traditional cellular networks, resulting in higher transmission rates. Additionally, the transmission rate of non-cooperative scenarios is slightly lower than that of cooperative scenarios due to independent learning and training. In Fig. 5(b), with the number of users fixed at 10, an increase in the number of APs leads to an increase in the uplink transmission rate. The rising number of APs mitigates the channel fading effect between serving users and APs. For UCMEC, the gain from AP collaborative transmission surpasses that of single AP service in CBO. Specifically, Proposed (Coop) at most achieves an increase in transmission rate of 12.88\% - 326.99\% compared to other schemes. In Fig. 5(c), with the number of users fixed at 10 and the number of APs fixed at 50, increasing the maximum transmission power results in a higher user uplink transmission rate. This is because the broader range of feasible solutions allows agents to make better power control decisions. Compared to other schemes, the transmission rate of Proposed (Coop) at most shows an increase of 8.13\% -227.94\%. Finally, in Fig. 5(d), when the AP cluster size increases, CBO does not obtain any gain due to the absence of AP clustering transmission. However, for UCMEC, enlarging the AP cluster size enhances the diversity gain and improves resistance to interference and fading. Specifically, Proposed (Coop) at most demonstrates an increase in transmission rate of 10.05\% - 343.56\%.

\subsubsection{Average Total Delay Performance}

In Fig. 6, we analyzed the long-term average total delay under different parameters. As shown in Fig. 6 (a), the total delay will increase as the number of users increases. In addition to the increase of uplink Transmission delay caused by the uplink transmission rate, the increase of offloading requests will also increase the load of fronthaul, increase the interference among APs, and increase the backhaul delay. In addition, new users will also compete for limited computing resources at the CPU, resulting in increased task processing delay. Compared to other schemes, the total delay of Proposed (Coop) can be at most decreased by 8.33\% - 35.06\%. As shown in Fig. 6 (b), as the number of APs increases, the total delay will decrease. In addition to reducing the uplink transmission delay, the increase of AP may also reduce the fronthaul transmission delay. Compared to other schemes, the total delay of Proposed (Coop) can be at most decreased by 19.59\% - 45.57\%. In Fig. 6 (c), when the maximum uplink transmit power increases, the uplink transmission rate will increase, reducing the uplink transmission delay. In addition, power control decisions can also affect the agent's decision on offloading, thus also affecting both fronthaul and task processing delay. However, when the maximum transmit power reaches above 0.2W, increasing the maximum transmission power does not bring more gain to delay performance. Compared to other schemes, the total delay of the Proposed (Coop) can at most be decreased by 16.74\% - 42.79\%. In Fig. 6 (d), as the AP cluster size increases, the delay of UCMEC will decrease. This is because the uplink Transmission delay will decrease significantly. However, the delay gain achieved by increasing the AP cluster size diminishes gradually due to the growing transmission requests from the AP to the CPU on the fronthaul, leading to increased AP interference and subsequent fronthaul delay. Specifically, compared to other scenarios, the total latency of the UCMEC in the cooperative scenario is at most reduced by 16.31\% - 38.24\%.

\subsubsection{Research Insights}

From the simulation results, we can summarize the following insights:
\begin{itemize}
    \item \textbf{Superiority of UCMEC in throughput and delay performance:} The AP cooperative transmission service can reduce the fading and interference of the access link, thus improving the uplink throughput and reducing the uplink transmission delay. In addition, the MmWave-based fronthaul ensures that the fronthaul overhead does not offset the gain of uplink transmission. The proposed decentralized resource allocation schemes can effectively improve the resource utilization efficiency in UCMEC, providing users with efficient transmission and computing services. Specifically, compared to traditional cellular-based MEC, UCMEC can at most achieve a 343.56\% increase in uplink transmission rate and reduce total delay by 45.57\%.
    
    \item \textbf{The trade-off of AP deployment:} Compared to cellular-based MEC, an increase in the number of APs and the size of AP clusters can bring significant transmission gain to UCMEC. However, the transmission gain brought by the number of APs and AP cluster size will gradually decrease. Therefore, considering the cost and energy consumption of AP, it is necessary to balance the AP density and the AP cluster size in actual AP deployment.

    \item \textbf{Surprising performance of non-cooperation schemes:} Collaborative cases enable agents to learn better strategies through global information, achieving superior performance compared to non-collaborative scenarios. However, in non-collaborative scenarios, shorter offline training time, easier online deployment, and lower information exchange frequency reduce actual signaling overhead. In the Proposed scheme, compared to the cooperative case, the non-cooperative case only at most experiences a 12.88\% decrease in uplink transmission rate and 19.59\% increase in total delay performance. This finding aligns with the conclusion in [1], where IL-based algorithms can achieve close performance to CTDE-based algorithms in some scenarios.
\end{itemize}

\section{Conclusion}
In this paper, we propose UCMEC, which combines the user-centric transmission services with the MEC computing services. We aim to minimize the long-term average latency of the system by jointly optimizing task offloading, power control, and computing resource allocation. To ensure efficient dynamic resource allocation, we propose two decentralized optimization schemes based on convex optimization and MADRL for cooperative and non- cooperative cases, respectively. The simulation results validate the effectiveness of UCMEC and the proposed optimization schemes. Specifically, UCMEC can improve uplink throughput by up to 343.56\% and reduce total latency by 45.57\% compared to traditional MEC.

\bibliographystyle{IEEEtran}
\bibliography{ref}	
\end{document}